# Journal Basket: Supporting Cross-Disciplinary Literature Searching and Methodological Transparency

Associate Professor Michael Cahalane | School of Information Systems and Technology Management, UNSW Business School, UNSW Sydney | m.cahalane@unsw.edu.au

## Abstract

Business students beginning a research project must learn how to identify relevant scholarly publications, distinguish disciplinary sources, construct database searches, and report the decisions behind their literature reviews. These tasks can be difficult because journal classification systems and bibliographic databases organize scholarly information differently. This paper introduces Journal Basket, a browser-based tool developed to help business students conduct journal-bounded literature searches and currently available in beta. Journal Basket enables users to construct journal collections from the Australian Business Deans Council (ABDC) Journal Quality List, choose from 42 business discipline presets, entire ABDC fields or the Financial Times FT50 list, apply journal rating thresholds, and generate search expressions for Scopus, Web of Science and Business Source. Users can also inspect and modify journal collections, download journal inventories, and generate methodological descriptions and deterministic identifiers for journal selections. Companion learning resources use explanations, examples and interactive activities to teach review methods, database choice, referencing and the use of contemporary AI tools. Drawing on the journal-basket concept used by Litbaskets.io, I describe the educational motivation, design rationale, technical architecture, and practical workflow of Journal Basket. I distinguish reproducing a journal collection from reproducing an entire database search and discuss the limitations of journal rankings as search boundaries. The paper addresses scholarly information retrieval through a workflow for defining journal boundaries, generating database-specific queries and documenting search decisions.

## 1. Introduction

An academic literature review begins with a basic question: which existing publications should a researcher read? Answering it requires more than entering keywords into a search box. Researchers decide which databases to consult, which terms represent their topic, which publication types to consider, and which sources fall inside the scope of their investigation. Those decisions shape the body of evidence that is eventually interpreted. Guidance in Information Systems (IS) research has long emphasized the need to describe and justify the literature search (Webster & Watson, 2002; vom Brocke et al., 2009).

For business students conducting an initial research project, the difficulty is partly conceptual. A student interested in artificial intelligence and employee decision-making may be able to articulate the topic but may not know which academic journals publish relevant work. The same phenomenon might be studied in Information Systems, management, organizational behavior, or accounting, using different terms and theoretical perspectives. A search of one journal group therefore captures a particular scholarly conversation, not necessarily the entire literature on the topic. A paper on AI-supported hiring might be framed as an information systems problem, a human resource management problem, or a question about fairness and organizational responsibility. Each framing may lead to different journals and different terminology. Students need a search process that helps them recognize these differences and choose sources deliberately.

The importance of looking beyond familiar IS journals is also recognized in research on the development of the discipline. Tarafdar and Davison (2018) argue that the growing presence of digital technologies in organizational and social life presents IS researchers with two connected tasks: advancing knowledge within IS and engaging with other disciplines to understand technology-related phenomena. In their analysis of 176 articles published in the AIS Senior Scholars' Basket of Eight between 2010 and 2014, contributions were predominantly intradisciplinary, with comparatively few classified as cross-disciplinary or interdisciplinary. This finding raises a question for search design: how well does a search confined to familiar IS outlets serve a question that spans several fields?

Journal Basket also grew out of more than a decade of teaching and supervising research in Information Systems. Across undergraduate and doctoral supervision, I have repeatedly explained how to move from a research question to a defensible literature search: where to look, how to decide which journals to include, and how to report those decisions. Students need guidance they can return to as their projects develop, rather than a set of instructions delivered once at the beginning. That guidance also needs to address students' use of generative AI and the decisions that require their own judgement. I developed Journal Basket in response to these recurring needs.

A journal-bounded search starts by deciding which journals to search. This selected group is called a journal basket, or simply a basket. Researchers then search for relevant articles within those journals. The basket sets the boundaries of the search; keywords determine which articles within those boundaries are retrieved.

Constructing a basket manually can be demanding. Journal names may change, print and online versions may have different International Standard Serial Numbers (ISSNs), and bibliographic databases use different commands for restricting a search to journals. A basket containing hundreds of journals can also produce queries too large to execute in one operation. Students may consequently spend considerable effort managing syntax before they have understood the methodological implications of their choices.

Journal Basket (https://www.journalbasket.com/) provides a reusable, guided resource for making these decisions. Its name acknowledges an earlier Information Systems tradition. The original Association for Information Systems (AIS) Senior Scholars' Basket of Eight identified eight leading IS journals. Boell and Wang (2019) drew on this basket terminology when creating LitBaskets (https://litbaskets.io/). Rather than limiting exploratory searches to those eight outlets, LitBaskets offered baskets of different sizes so researchers could search more widely across journals relevant to IS. Journal Basket takes inspiration from that approach but uses the Australian Business Deans Council (ABDC) Journal Quality List to help users build and combine journal baskets across business disciplines. The ABDC list gives students a recognizable framework for identifying business publication outlets, especially when a research question crosses several fields. Its coverage and ratings are discussed in Section 2.2.

This paper describes a practical research tool and explains the methodological choices embedded in its design. Its contribution is an ABDC-based workflow for combining journal collections across business disciplines, translating those selections into searches for three databases, and retaining an inspectable record of the selected journals and settings. I explain the accompanying teaching resources and document a dated search execution. Section 5.3 sets out the evaluation needed to assess educational and research benefits.

The paper is structured as follows. Section 2 introduces the methodological choices involved in journal-bounded searching, including exploratory searching, the ABDC list, and interdisciplinary scope. Section 3 explains Journal Basket's user workflow, compares its design with LitBaskets, develops a worked student example, and explains the companion learning resources. Section 4 describes the technical implementation, from journal data preparation to database query generation and search documentation. Section 5 considers applications, limitations, and future evaluation, before Section 6 concludes.

## 2. Background and Methodological Motivation

### 2.1. From a research question to search terms

A research question rarely maps onto a single expression used consistently across published articles. A student investigating AI-assisted decision-making may encounter terms such as “algorithmic decision-making,” “decision support systems,” and “human–AI collaboration.” Search design therefore involves identifying alternative expressions for the same or related concepts. In a bibliographic database, Boolean operators combine these expressions: OR joins alternatives, while AND requires two concept groups to occur together. For example, (“artificial intelligence” OR “machine learning”) AND (“decision making” OR “decision support”) provides a starting expression to refine against the research question and initial results.

Literature reviews differ in purpose and design. A systematic review seeks to identify evidence through an explicit protocol, whereas other review types may prioritize conceptual integration or mapping an emerging field. The search strategy should follow the review's purpose rather than being selected merely because a tool makes it convenient (Snyder, 2019; Paré et al., 2015).

Bibliographic databases such as Scopus, Web of Science and Business Source provide structured searches across indexed scholarly records. Their advanced search interfaces permit restrictions on fields, publication dates, document types, and source identifiers. These facilities allow researchers to specify and report their search logic using database-specific syntax.

The choice of database also influences which publications can be found. Evidence from systematic review research shows that databases vary in coverage and that combinations of databases can retrieve records missed by a single source (Bramer et al., 2017). Although those findings arise largely outside business research, the underlying methodological point applies: a well-constructed journal basket cannot compensate for publications that a chosen database does not index.

Journal Basket translates the user's journal selection and chosen terms into the syntax of the selected database. The researcher develops the substantive search strategy, including the concepts and synonyms; the software constructs the source restrictions.

Not all literature searches begin with a settled vocabulary or a fixed list of sources. During exploratory searching, researchers read early results, encounter unfamiliar concepts, and revise both their search terms and the journals they consult. Wang et al. (2026) describe this as an iterative process in which searching and reading develop together. This is especially relevant to students beginning a research project: an initial search can help them discover how a topic is discussed, while later searches can become more focused as their understanding improves. A journal basket can provide a useful starting boundary, provided that researchers remain willing to reconsider it.

### 2.2. Why use a journal basket?

A journal basket is useful when the question concerns research published within a particular scholarly community. For example, a student might ask how leading Information Systems journals have discussed algorithmic management. Selecting those journals creates a clear boundary that the student can explain to a reader. The AIS Senior Scholars' Basket of Eight offers a familiar example of such a bounded set. Boell and Wang (2019) recognized that relevant IS work also appears beyond those eight journals. Their LitBaskets tool (https://litbaskets.io/) allows researchers to choose progressively larger groups of journals, starting with the Basket of Eight and expanding outward.

The same restriction would be less defensible if the student intended to review all empirical evidence on AI adoption in professional services. Relevant studies may appear in management, psychology, computer science, conference proceedings, and journals outside any selected basket. A journal-bounded search is therefore a purposeful sampling strategy for published scholarship, not a universal substitute for comprehensive searching. Its adequacy must be assessed against the review objective (Okoli, 2015; Webster & Watson, 2002).

Many business questions cross disciplinary lines. Research on algorithmic management, for instance, may appear in Information Systems, management, human resource management, and organizational studies journals. Searching only one of these groups can miss perspectives developed elsewhere. A practical advantage of Journal Basket is that users can select several disciplines and bring their journals together in one basket, with duplicate entries removed. This makes it easier to explore work across fields without constructing several separate journal lists by hand. Researchers can then inspect which publication outlets become eligible when the search extends beyond its original discipline. For example, adding marketing journals to an Information Systems basket may bring consumer behavior and trust into view, while adding management journals may foreground organizational strategy and work practices. The expanded list gives researchers a concrete basis for judging whether those additional perspectives are relevant.

The Australian Business Deans Council (ABDC) Journal Quality List offers a useful starting point for this task. It brings together business journals, assigns them to Fields of Research, and groups them into A*, A, B, and C rating categories. For a student who does not yet know the journals in accounting, marketing, or Information Systems, the list provides a way to discover established publication outlets. It also gives researchers a shared reference for explaining why certain journals were considered. The ratings describe journals, not the quality or relevance of every article they publish. A highly relevant study can appear outside an A* journal, and a review should not equate journal rank with evidence quality (DORA, 2013).

Journal Basket uses the ABDC list as a common starting framework for journal selection across business disciplines. Users choose a discipline and rating categories, inspect the journals included, and may combine disciplines where their research question spans several fields. For example, limiting a basket to A* and A journals may suit a review explicitly focused on highly rated business outlets, with the restriction justified by that focus. The website makes clear that Journal Basket is independent of, and not endorsed by, the ABDC.

### 2.3. Searching across disciplines and developing interdisciplinary knowledge

Searching journals from several disciplines creates opportunities for interdisciplinary research. A student examining AI-assisted employee decisions might search IS, management, and organizational behavior journals. Comparing how these fields understand decision-making and bringing their ideas into a coherent account, requires further interpretive work.

Tarafdar and Davison (2018) distinguish contributions developed primarily within IS from those that draw on other disciplines, contribute to multiple fields, or integrate concepts from different disciplines to develop new explanations. A study can consult psychology or management research and still make its principal contribution to IS. An interdisciplinary contribution, as they define it, involves integrating ideas from different fields. For students, the practical lesson is that a broader search creates opportunities to encounter different perspectives, but the work of comparing and integrating those perspectives remains their responsibility.

Journal Basket supports the search stage of that process. Users can begin with an IS-focused basket, add journals from related business disciplines, and inspect which outlets the broader selection includes. This makes the publication boundaries of a cross-disciplinary search easier to construct and report.

### 2.4. Teaching the distinction between search boundaries and search content

For a student new to research, it helps to separate two questions: “Where should I look?” and “What should I look for?” The basket answers the first question by identifying journals. The keywords answer the second by identifying topics and concepts within those journals.

Journal Basket places these choices in separate steps so students can see what changes when they add a discipline, include more rating categories, or revise a search phrase.

A student studying the organizational use of generative AI might initially select Information Systems journals and subsequently add management and marketing. The expanded basket can reveal additional publication outlets without requiring the student to know their names in advance. This is especially relevant when research questions span technological, organizational and market perspectives. Students must still justify the selected disciplines and consider scholarship outside the ABDC list, including relevant conference proceedings, books and journals classified elsewhere.

## 3. Software Overview and Design Rationale

### 3.1. User workflow

Journal Basket is organized into seven pages, which its menu presents in two groups. Three pages support the search workflow: Search, Journals, and Write-up. Four are guides: Reviews (conducting a literature review), AI (using generative AI tools in research), Citing (citing and referencing), and Help. Search guides the user through entering search words, choosing one or more disciplines, and selecting journal ratings. Journals shows the resulting basket so users can check or adjust its contents. Write-up helps users record what they searched, and Help explains unfamiliar terms and procedures. The sequence supports a basic research workflow: define the topic and search boundaries, check the journal list, and document the choices made.

Figure 1 illustrates the Search page before a discipline has been selected. It presents four numbered steps: typing search words, choosing a discipline, choosing how many journals to search by ABDC rating, and choosing a database. Further options, including publication years, document type, and title-only searching, are available beneath the search box, and a narrated video walkthrough is linked above the steps. The rating options are described in plain language alongside their classification labels, with A* and A marked as recommended and single ratings offered separately. For users working across disciplines, two shortcuts sit beneath the discipline menu: the A* journals of every ABDC field (219 journals) and the 50 journals of the Financial Times FT50 list, marked as new in Figure 1. Step 4 offers searches in Scopus, Web of Science and Business Source, which open through a library's login page (UNSW Library by default, changeable in the Help page's settings). Journal counts appear once the user has chosen a discipline.

The screen is designed around decisions a student can recognize, rather than asking them to write database commands from the outset. Choosing disciplines defines where to search; choosing ratings adjusts the breadth of the journal list; entering keywords defines the topic. Once a discipline is chosen, step 4 states how many journals will be searched and at which ratings, and the Journals page lists them. These visible choices give students and supervisors a basis for discussing the search strategy.

The software supports 42 discipline presets, each of the 17 ABDC Fields of Research on its own, a selection spanning all fields, and the 50 journals of the Financial Times FT50 list (2026 edition; Financial Times, 2026), all of which appear on the ABDC list. Multiple selected disciplines are combined into one basket with duplicate journals removed. The discipline presets are constructed from ABDC field membership and curated specialist additions; they remain drafts awaiting expert review. For a discipline with specialist additions, users can search its home field, its specialist journals, or both. Researchers should inspect the journals in a preset to assess how well it fits their question.

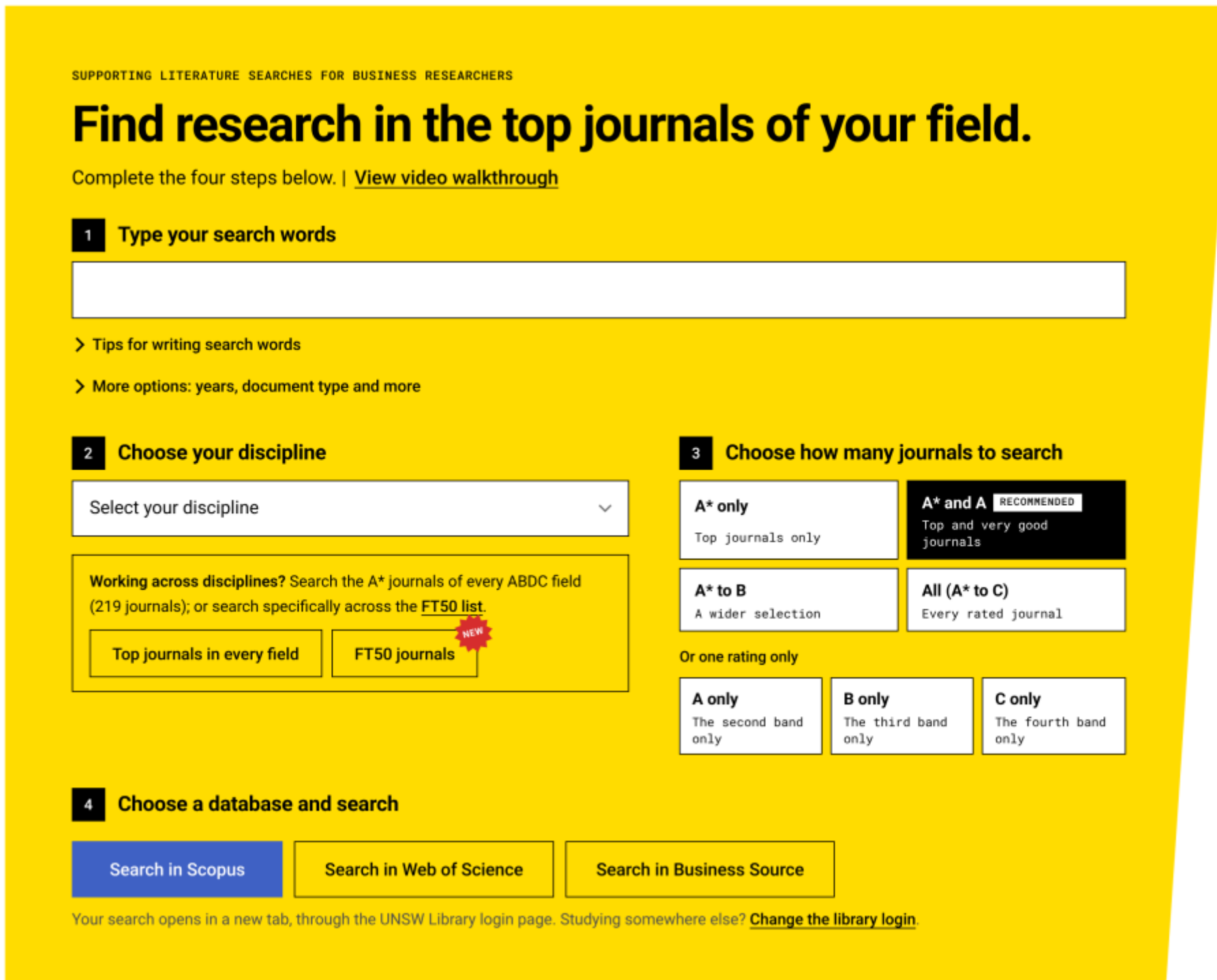

***Figure 1. Journal Basket's Search page (version 0.30.3) showing the four-step search construction workflow and the cross-disciplinary shortcuts before a discipline is chosen.***

Combining disciplines is one of the tool's main practical benefits. A student researching digital platforms and consumer privacy, for example, could bring together journals in Information Systems, marketing, and management. This can help users encounter conversations that would be less visible from a single disciplinary starting point.

Journal Basket does not store a user's search choices or show them in the browser's address bar, in keeping with its privacy design (Section 4.1). Users can share a proposed basket with a supervisor through the downloadable journal list and journal list code (Section 4.4). A dated journal inventory matters most when the exact selection must be preserved, since underlying source data or software behavior may change in later releases.

### 3.2. Relationship to LitBaskets: design and functionality

LitBaskets has been part of my guidance to Information Systems students in recent years. It draws on 11 journal lists associated with IS to organize 847 journals into eight progressively larger baskets, starting with the original AIS Senior Scholars' Basket of Eight (Wang et al., 2026). Users can move between basket sizes, include or exclude particular journals, and export and import their configurations. Journal Basket builds on that flexible approach but changes the starting point: users select business disciplines and ABDC rating categories, then combine the journal collections. This responds to a recurring need in my supervision for searches that extend beyond an initially familiar set of IS journals. The tools also differ in their search destinations and documentation facilities. LitBaskets is centered on generating searches for Scopus; its published account describes a search panel, a basket-size slider, and a journal-selection interface (Wang et al., 2026). Journal Basket generates source-restricted searches for Scopus, Web of Science and Business Source and offers a Write-up area for recording selected journals and search settings.

The tools assemble their journal sets differently. LitBaskets combines source lists associated with IS and groups journals according to how often they occur across those lists (Wang et al., 2026). Journal Basket instead uses ABDC records, disciplinary presets, journal ratings, and user exclusions to construct its set. Its implementation validates journal identifiers during data preparation, converts the selected journals into source restrictions (Scopus source identifiers or ISSNs for Scopus, IS= for Web of Science, and the IS field code for Business Source), and divides long queries into executable components.

Table 1 summarizes these differences in journal selection, supported databases, and search documentation. Both tools allow users to select and refine journal collections independently of their search terms. Researchers examining established conversations within Information Systems may find LitBaskets' progressively broader IS-oriented collections useful. Those investigating questions that span disciplines, such as accounting, management, and Information Systems, can use Journal Basket to combine the corresponding journal collections directly.

***Table 1. Comparison of LitBaskets and Journal Basket functionality.***

| **Dimension** | **www.LitBaskets.io** | **www.JournalBasket.com** |
| --- | --- | --- |
| Primary orientation | Exploratory search in IS and adjacent areas | Journal-bounded search across business disciplines |
| Journal source | 11 IS-related lists; 847 journals | ABDC Journal Quality List |
| Initial selection | One of eight progressively larger IS-oriented baskets | Business disciplines, ABDC fields or the FT50 list, with journal rating categories |
| Broaden search | Choose a larger basket or customize journals | Add disciplines, expand rating categories, or customize journals |
| Journal customization | Include/exclude individual journals | Include/exclude individual journals |
| Search destination | Scopus | Scopus, Web of Science and Business Source |
| Reuse / reporting | Export and import journal configurations | Download journal inventories, methods text with references, and journal list codes |

### 3.3. A worked student example

Consider a student beginning a literature review with the question, “How is generative artificial intelligence being studied in auditing?” At this stage the question is exploratory: the student wants to learn which issues scholars have examined before narrowing the focus. An unrestricted search may return work from computer science, accounting, law, management, and other fields. The student uses Journal Basket to define an initial set of journals in which to look. This hypothetical example illustrates the student workflow.

In step 2 of the Search page (Figure 1), the student selects an auditing-related discipline, such as the Auditing preset. This creates a basket: a list of journals that will define the publication boundaries of the search. The student opens the Journals page to inspect the list and check whether it fits the research question. If the student is interested in how auditing scholars discuss AI, the initial basket may be appropriate. If the interest extends to how audit firms adopt AI-enabled systems, adding Information Systems and accounting disciplines may expose related work published outside specialist auditing outlets.

Combining disciplines is a deliberate change to the research scope. Journal Basket joins the selected journal lists and removes duplicate journal entries, so a journal appearing in more than one selected discipline is searched once. The student can compare a narrower auditing basket with the combined basket and explain which better fits the research question. The larger basket may return more material to screen. Comparing the baskets also helps the student clarify the scope: is the aim to understand how auditing journals discuss generative AI, or to understand the phenomenon across auditing, accounting, and Information Systems? The basket defines eligible publication outlets; search terms and subsequent screening determine which articles are retained.

In step 3, the student chooses which ABDC rating categories to include. Selecting A* and A journals produces a narrower list than including A*, A, B, and C. For an initial orientation exercise, a narrower selection may be manageable, but it cannot be justified simply by assuming that articles in higher-rated journals are individually better. The student should state the purpose of the restriction and, if the review aims for broader coverage, consider expanding the basket or supplementing it with other sources.

The topic terms go in step 1, at the top of the page, and the student can return to them at any point. A first attempt might use (“generative artificial intelligence” OR “large language model*”) AND (audit* OR assurance). Here OR collects alternative expressions for a concept, AND links the AI and auditing concepts, and the asterisk is a truncation symbol in databases that support it, potentially matching variants of a word. The student should check the precise syntax in the chosen database and refine the expression after inspecting initial results. Before opening a search, Journal Basket points out common typing mistakes, such as an unmatched quotation mark or bracket.

In step 4, the student selects Scopus, Web of Science or Business Source. Journal Basket constructs journal restrictions from the selected basket and combines them with the search terms in the syntax required by the chosen database. If the basket is large, the tool provides smaller search components and instructions for combining them. The student executes the search in the database, checks that it runs as intended, and records the complete expression, date, filters, and number of results. Article retrieval takes place in the bibliographic database.

Finally, the student uses the Write-up area to download the journal inventory and obtain a draft description of the basket. The inventory documents which journals were eligible to be searched, while the journal list code helps distinguish this selection from a different one. The student still needs to explain why the basket was chosen, remove duplicate article records where necessary, screen retrieved publications against inclusion criteria, and synthesize the evidence. A supervisor could ask the student to compare the two baskets and defend the final choice. The student can now describe the search procedure: which disciplines were selected, which ABDC ratings were included, whether any journals were removed, which terms were searched, and when and where the search was run. If the student later broadens the basket, the saved inventory makes the change visible. The record identifies the choices that shaped the evidence considered. Appendix A provides a separate execution record from software verification, including the complete query and journal inventory for a two-journal Scopus search.

### 3.4. Learning resources for research students

Journal Basket includes three learning resources: Literature Reviews in Business Research, Using AI Tools, and Citing and Referencing. I present the material through short explanations, worked examples, comparison tables and interactive activities that students can revisit while developing a review. The guides connect choices about searching with the reading and writing that follow.

The literature review guide compares Scopus, Web of Science, Business Source and Google Scholar in terms of coverage, search facilities and reproducibility. It introduces several review approaches, including narrative, integrative, systematic, scoping and bibliometric reviews, and helps students relate each to the purpose of their project. It also presents two complementary ways of working with the literature: Webster and Watson’s (2002) concept-centered synthesis, supported by a concept-matrix exercise, and Boell and Cecez-Kecmanovic’s (2014) hermeneutic approach, in which searching, reading and interpretation develop iteratively. Activities and examples help students move from collecting papers to comparing concepts and building an argument.

The AI guide recognizes that today’s research students will want to incorporate contemporary AI tools into their search and review process. It explains how general-purpose assistants such as ChatGPT and Claude, scholarly search tools and connectors, and database-integrated AI can assist with refining questions, developing search terms, discovering literature and organizing reading. Students are encouraged to test suggested terms in bibliographic databases, read the underlying papers and verify references and claims. The guide also addresses recording AI use and following institutional requirements.

The citing and referencing guide bring these activities into academic writing. Step-by-step examples, practice activities and short videos cover paraphrasing, concept-centered synthesis, citation tracing, reference management and DOI checks. The emphasis is on connecting an argument to its sources and checking references, as well as applying a citation style.

## 4. Technical Architecture and Data Processing

### 4.1. Browser-based implementation

Journal Basket is implemented primarily as a static client-side application using HTML, CSS, and JavaScript. Its user interface and search construction operate in the browser without requiring an account. The project is deployed on Vercel. A server endpoint records aggregate counts in an Upstash Redis store for searches directed to Scopus, Web of Science and Business Source, and for searches using the two cross-disciplinary shortcuts. The Help page displays these totals. The project documentation states that the endpoint does not record search terms or user identities.

The site also uses consent-controlled Google Analytics. Its implementation is designed to avoid transmitting search terms in analytics requests, and analytics are not loaded unless the visitor accepts the relevant cookies. These privacy provisions matter for student use because a search query can disclose an unpublished research interest.

### 4.2. Journal records and identifier quality

The version described here uses the 2025 ABDC Journal Quality List, version 3 dated 21 September 2026 (Australian Business Deans Council, 2026). The generated data-quality report records 2,648 journals. One entry, ISEAS Current Economic Affairs Series, has neither a usable ISSN nor a Scopus source identifier; it remains in the journal inventory but is omitted from generated searches, with a notice to the user. The software converts the source spreadsheet into structured journal data and maintains a separate corrections file. Additional ISSNs, including those associated with former titles, are recorded for 58 journals using evidence from Crossref, Scopus, the ISSN Portal and Web of Science. A total of 2,095 journals are matched to Scopus source identifiers. Keeping corrections separate preserves the original classification data and makes the changes available for inspection.

An International Standard Serial Number (ISSN) is a code used to identify a journal or other continuing publication. Some journals have one ISSN for the print edition and another for the online edition. A journal title can also change over time. These details matter because Scopus, Web of Science and Business Source use journal identifiers when applying source restrictions. If an identifier is missing or incorrect, a search may leave out articles from a journal that the researcher intended to include. Journal Basket therefore checks and records journal identifier information as part of preparing its data.

The data-quality report provides concrete examples of why validation matters. It records corrections to malformed or missing identifiers and cases where adding a corroborated identifier changes the number of indexed records retrievable in Scopus. These cases show how identifier handling can affect retrieval coverage.

### 4.3. Generating executable searches

After a user selects a basket, Journal Basket converts the journal list into database-specific source restrictions. Each database identifies journals in its own way. For Scopus, the software uses SRCID() for the 2,095 journals matched to source identifiers and ISSN() for the remaining journals with usable ISSNs. Web of Science uses IS=, and Business

Source uses the IS field code. Journal Basket handles this translation, so users do not need to type hundreds of journal codes themselves.

Large baskets create a practical problem because a database interface may reject an excessively long expression. By default, each Scopus or Web of Science query part holds up to 100 journal identifiers, adjustable to as few as 20 in the Help settings. A Business Source part holds up to 40. These limits count identifiers, so a journal with several ISSNs or source identifiers occupies more than one place. Web of Science parts are also constrained by a 2,250-character link limit to accommodate its sign-in page. Journal Basket divides large selections into parts and provides instructions for combining them with OR in Scopus and Web of Science, while Business Source parts are run one after another. For Scopus, the whole search can instead be copied as a single expression for its Advanced Search. Researchers should check each executed search and its combined results in the target database.

The all-fields option contains 219 A* journals, 849 A* or A journals, 1,708 A*, A or B journals, and 2,648 journals across all ratings. With short search terms and the default settings, the 2,647 searchable journals require 29 Scopus parts, 47 Web of Science parts and 119 Business Source parts, while the 219 A* journals require 3, 5 and 11 parts, respectively. These figures apply to the dataset and settings in version 0.30.3.

### 4.4. Documenting a basket

The Write-up page generates a draft methods paragraph from the user's selected database, journal selection criteria, search terms, and search settings. As illustrated in Figure 2, the generated paragraph identifies the ABDC Journal Quality List edition, disciplinary classification, journal ratings, number of selected journals, journal identifiers used to construct the search, the Journal Basket version, and the journal list code. Where a search must be divided into multiple queries because of its length, the paragraph also describes how these searches are combined. Researchers can copy the generated text into their manuscript and adapt it to reflect their actual search procedures. Below the paragraph, the Write-up page lists the references it relies on in APA style (the ABDC list, and the Financial Times' list when the FT50 is searched) and provides a prompt that users can paste into their own AI assistant to help check the paragraph.

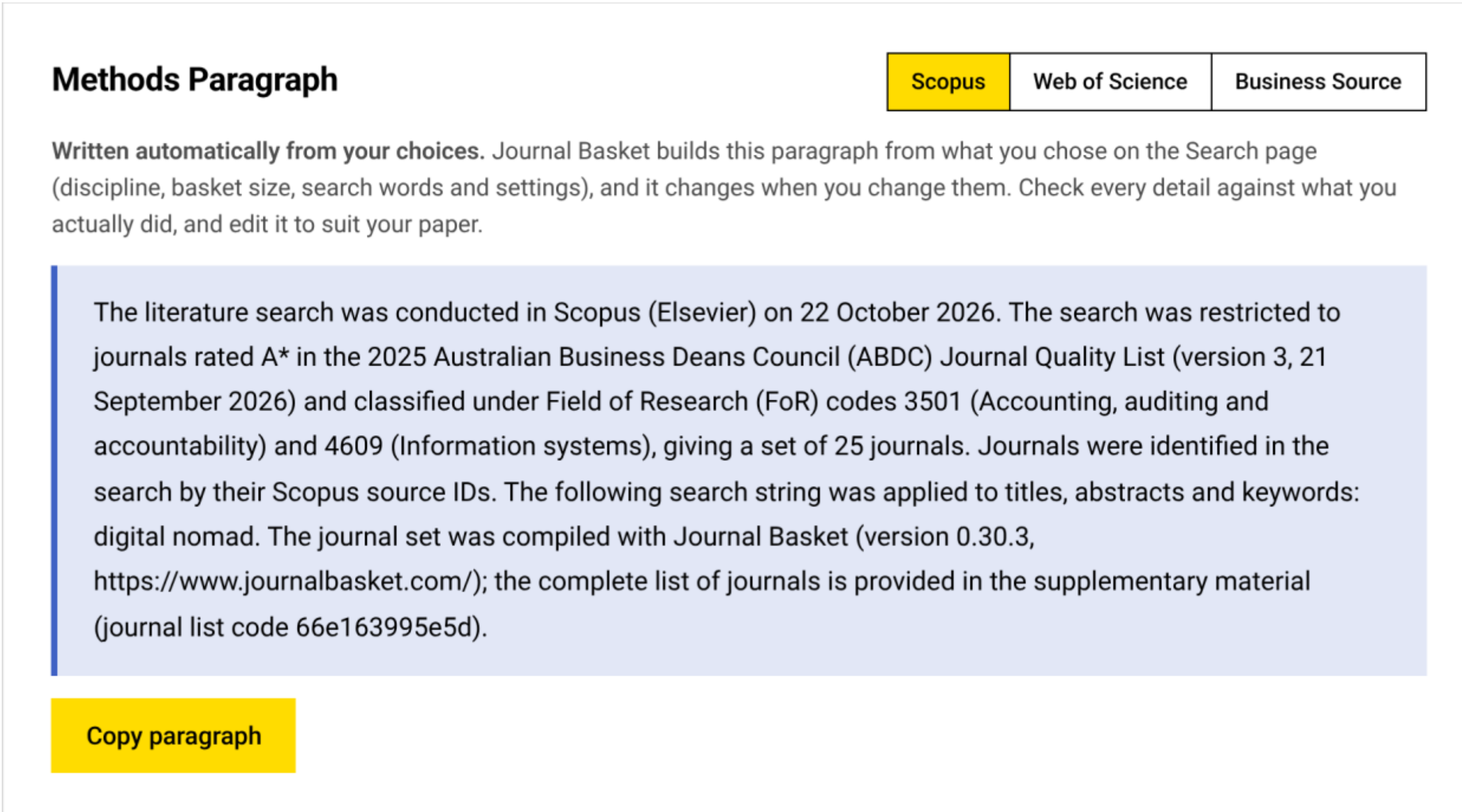

***Figure 2. Automatically generated methods paragraph for a journal-bounded literature search in Journal Basket version 0.30.3.***

The downloadable journal inventory is accompanied by a deterministic journal list code. The code is deterministic because the same inputs produce the same code. In Journal Basket, this code is the first 12 characters of a SHA-256 fingerprint of the ABDC list's edition and version and the sorted journal identifiers, providing a compact reference for comparing journal collections. For example, two researchers examining artificial intelligence in Information Systems might use identical journal selections but different search terms. Their journal list codes would match, even though the searches themselves could retrieve different publications. The code refers to the journal selection and ABDC list version; it does not encode the search terms.

Reproducing a literature search requires enough information for another researcher to repeat the procedure and assess the evidence. Established literature review guidance emphasizes the importance of explicitly documenting information sources, search strategies, and selection procedures (Okoli, 2015). Reproducing a complete literature search also requires the search expressions, bibliographic database, execution date, applied filters, and subsequent screening decisions. Even when these details are preserved, differences in database coverage or indexing over time may affect the results obtained.

The generated methods paragraph reduces the need to reconstruct the search settings manually. Researchers should check it against the searches actually conducted and explain why the selected journal boundaries fit their questions. For students, the distinction is useful: documenting a search establishes what was done, whereas methodological justification explains why those decisions were appropriate.

### 4.5. Development and verification

The project includes build scripts, automated tests, a changelog, decision records, accessibility checks, and data-quality documentation. The development report for version 0.30.3 (22 October 2026) records 115 passing automated tests. The preceding version 0.30.2 report (21 October 2026) records no findings from an automated WCAG 2.1 AA accessibility check across 72 page views and successful browser checks of 427 combinations of journal selection and rating level.

A separate verification run on 21 October 2026 at 20:28 UTC checked 24 generated searches against the live site running version 0.30.1. Sixteen cases were executed through Elsevier's Scopus Search API or Clarivate's Web of Science Starter API; eight Business Source cases were checked for link construction only. The executed query parts returned 12,458 records, of which 612 were checked for journal membership and any specified year and document-type restrictions. All 24 cases passed their respective checks. Appendix A records one complete execution from this run. A monthly health-check workflow has also been configured to repeat search and source-list checks. These records document implementation checks; evaluation of usability and methodological effectiveness is proposed in Section 5.3.

## 5. Applications, Implications, and Limitations

### 5.1. Student learning and supervision

For students, Journal Basket can turn an abstract discussion about search strategy into a set of choices they can inspect. This design reflects a recurring issue in my supervision: students can often explain their research topic but struggle to justify why their search should stop at the boundaries of Information Systems journals. I have found it useful to ask them to compare an IS-only basket with one that also includes management or marketing outlets, and to explain what each selection would allow them to see or potentially miss. Journal Basket makes this comparison practical. Students can inspect the journals added when another discipline is selected and discuss whether that wider scope is warranted by the research question. The exercise makes the trade-off between focus and breadth concrete.

The educational purpose extends to research writing. In supervising undergraduate and doctoral work, I have often encountered uncertainty about how to describe and justify a literature search in a methodology section. Journal Basket responds by giving students a journal inventory and a draft description of their settings, which they can examine alongside the search they actually performed. These outputs allow a supervisor to ask why particular disciplines and ratings were selected, what sources were excluded, and what supplementary searches might be necessary. The learning resources described in Section 3.4 provide examples and exercises for developing these explanations.

### 5.2. Use by experienced researchers

The principle of adjustable journal boundaries extends beyond teaching. Wang et al. (2026) describe exploratory searches as iterative: researchers may broaden or narrow the journals consulted as they learn more about a topic. Journal Basket supports a related practice by allowing an experienced researcher to move from a single business discipline to a combined basket and retain a record of that change.

Researchers with more experience may also find the tool useful when they need to search a clearly defined set of business journals. Instead of assembling and checking a long list of journal identifiers manually, they can start from an ABDC-based selection, adjust it, and export a record of the journals searched.

For interdisciplinary reviews, the tool can support a deliberate expansion strategy. A researcher may begin with a core discipline, inspect the initial basket, and add adjacent fields when the research question calls for perspectives developed elsewhere. For example, work on digital technologies may require literature from organisational behaviour, marketing, or accounting alongside IS. Tarafdar and Davison (2018) distinguish consulting scholarship from other disciplines from integrating its concepts into a substantive knowledge contribution. Journal Basket helps researchers organise the search across disciplines. The documented basket provides a basis for explaining why particular disciplines were included and what the search may still omit.

Journal Basket does not assess article quality, resolve theoretical disagreements, perform screening, or synthesise findings. Nor does it replace searches of other databases, conference proceedings, books, or grey literature where these sources are relevant.

### 5.3. Limitations and future evaluation

Three limitations affect the use of journal baskets. First, a classification-based basket reflects the coverage and disciplinary assumptions of the source list and the project's curated presets. Second, journal ratings apply to outlets, not to the relevance or quality of individual articles. Third, database indexing and identifier histories may prevent an apparently correct query from retrieving every potentially relevant record.

A related limitation is that a journal's disciplinary classification cannot establish the intellectual orientation of every article published within it. Tarafdar and Davison (2018) explicitly acknowledge that their Basket of Eight analysis did not capture IS-related scholarship published in non-IS journals and identify the value of examining outlets in other disciplines. A combined ABDC-based basket may still miss relevant scholarship, particularly in fields beyond business. Journal selection must therefore be justified in relation to the review question and supplemented where necessary.

The checks reported in Section 4.5 test whether generated searches are accepted by Scopus and Web of Science and whether returned records belong to the selected journals. They do not measure retrieval completeness. Future evaluation should examine whether generated searches match independently verified reference searches across representative baskets and all three supported databases. A separate usability study could observe novice business students constructing and explaining searches, comparing their decisions with and without the tool. Such studies could test error rates, comprehension, workload, and the quality of methodological justification. Until that evidence exists, educational and productivity benefits remain intended outcomes rather than demonstrated effects.

## 6. Conclusion

Journal Basket aims to help business students and researchers build and explain journal-bounded literature searches. Building on the AIS Basket of Eight tradition and the broader exploratory approach of LitBaskets, Journal Basket uses the ABDC Journal Quality List to make basket selection possible across business disciplines. Users can combine disciplines, generate searches for Scopus, Web of Science and Business Source, and retain a record of their journal choices. The workflow supports scholarly information retrieval by linking explicit journal boundaries to database-specific queries and a record of the search decisions. The combination is particularly relevant to business students who are still learning how a literature review is designed. For experienced researchers, the same workflow offers a convenient way to build and document searches that draw on several business disciplines.

## Software Availability

Journal Basket is available as a beta application at https://www.journalbasket.com/. This paper describes version 0.30.3 (22 October 2026). The three learning resources are linked in Section 3.4, and Appendix A records the query and journal inventory for the executed example. Journal Basket is independent of, and not endorsed by, the Australian Business Deans Council or the Financial Times.

## Declaration of Generative AI Use

I used OpenAI's ChatGPT and Anthropic's Claude to assist with drafting and editing this manuscript. This assistance included organizing the manuscript, improving readability, and developing explanations of the software's functionality and methodological foundations. I reviewed and revised the AI-assisted content and retain full responsibility for the manuscript's accuracy, scholarly integrity, interpretations, and conclusions.

## Appendix A. Recorded search execution

The project's verification run dated 21 October 2026 at 20:28 UTC (Journal Basket version 0.30.1; seed 762933, case 13) records the following Scopus search. The basket comprised the two A-rated specialist journals in the Management Accounting preset. The phrase "corporate governance" was searched in titles, abstracts and keywords, with no publication-year or document-type restriction. The query was executed through the Scopus Search API and returned 14 records; all 14 were checked for membership of the selected journals. This is an implementation example rather than a completed literature review.

Complete search expression:

```
TITLE-ABS-KEY("corporate governance") AND (SRCID(4100151511) OR
SRCID(19900192221))
```

***Table A1. Complete journal inventory for the recorded search.***

| Journal | ABDC rating | ISSNs | Scopus source ID |
|---|---|---|---|
| Advances in Management Accounting | A | 1474-7871 | 4100151511 |
| Journal of Management Control | A | 2191-4761; 2191-477X | 19900192221 |

**Journal list code:** b8872e8ee046. **Source classification:** 2025 ABDC Journal Quality List, version 3 (21 September 2026). The inventory is complete for this example; the query uses Scopus source identifiers rather than ISSNs. The count records the result at the execution date and may change as Scopus is updated.